\documentclass[11pt]{article}

\usepackage[preprint]{acl}

\usepackage{times}
\usepackage{latexsym}

\usepackage[T1]{fontenc}

\usepackage[utf8]{inputenc}

\usepackage{microtype}

\usepackage{inconsolata}

\usepackage{graphicx}

\usepackage{amsmath}
\usepackage{amsfonts}
\usepackage{booktabs}
\usepackage{multirow}
\usepackage{enumitem}
\usepackage[table]{xcolor}
\definecolor{myteal}{HTML}{A4DDD3}

\usepackage{tcolorbox}
\tcbuselibrary{skins,breakable}
\definecolor{agentQ}{HTML}{D6EAF8}
\definecolor{agentE}{HTML}{D5F5E3}
\definecolor{agentG}{HTML}{FDEBD0}
\definecolor{agentQtitle}{HTML}{2E86C1}
\definecolor{agentEtitle}{HTML}{1E8449}
\definecolor{agentGtitle}{HTML}{CA6F1E}

\title{Hallucination-Aware Multi-Agent RAG via Internal Attention Rewards}

\author{
\textbf{Runsong Jia}\textsuperscript{1} \quad
\textbf{Qian Chang}\textsuperscript{2} \quad
\textbf{Mengjia Wu}\textsuperscript{1} \\
\textbf{Ciprian Doru Giurc\u{a}neanu}\textsuperscript{2} \quad
\textbf{Jie Lu}\textsuperscript{1} \quad
\textbf{Yi Zhang}\textsuperscript{1}\thanks{Corresponding author.} \\
\textsuperscript{1} University of Technology Sydney, Sydney, Australia \\
\textsuperscript{2} University of Auckland, Auckland, New Zealand \\
\texttt{\{runsong.jia, mengjia.wu, jie.lu, yi.zhang\}@uts.edu.au}, \\
\texttt{qcha783@aucklanduni.ac.nz, c.giurcaneanu@auckland.ac.nz}
}

\begin{document}
\maketitle
\begin{abstract}
Retrieval-augmented generation (RAG) grounds language models in retrieved evidence, but its outputs often contain claims unsupported by that evidence. Recent multi-agent RAG systems decompose the pipeline into cooperative decisions, yet they typically optimize all agents under a shared task reward such as EM or F1, which measures answer overlap rather than faithfulness to evidence. We propose a hallucination-aware multi-agent RAG framework that combines the task reward with an internal hallucination-risk signal computed from the generator's attention-head norms. Three role-specific LoRA adapters, a query planner, an evidence selector, and an answer generator, share a frozen backbone. A lightweight critic is trained once on TruthfulQA truthful/false pairs and then kept frozen, providing a complementary risk score that is added to the shared reward during cooperative PPO training. Across three backbones (Qwen3-8B, Llama3-8B, Llama2-13B) and three benchmarks (RAGTruth, HotpotQA, 2WikiMultiHopQA), our method improves task performance and reduces hallucination rates over the strongest multi-agent baseline. Ablations and critic diagnostics show that the TruthfulQA-calibrated signal provides useful reward shaping that transfers across RAG tasks.
\end{abstract}

\section{Introduction}
Retrieval-augmented generation (RAG) grounds language models in external evidence rather than parametric memory~\cite{lewis2020rag,gao2024rag_survey}, with the goal of producing answers that are easier to verify. In practice, however, RAG systems still produce claims that the evidence does not support: an answer may name the correct entity while adding relations, dates, or numbers that the retrieved passages never mention~\cite{ji2023hallucination_survey}. These failures do not arise from a single stage of the pipeline. Ambiguous questions, noisy or partially relevant passages, and the generator itself can each lead to claims that go beyond the evidence~\cite{niu2024ragtruth}, and a single end-to-end objective does not indicate which stage is responsible.

Multi-agent RAG responds to this multi-stage nature by decomposing the pipeline into cooperative, learnable stages~\cite{chen2026mmoa}. A query planner makes the information need explicit, an evidence selector decides what to condition on, and an answer generator produces the final response. With this decomposition, the key design choice is no longer the architecture but the reward. When the three agents are optimized under a shared task reward such as EM or F1, hitting the reference span is enough to obtain most of the reward, while additional claims that go beyond the evidence are barely penalized. The reward signal therefore shapes the agents toward answer overlap, not toward grounding each claim in the evidence.

\begin{figure}[t]
\centering
\includegraphics[width=\columnwidth]{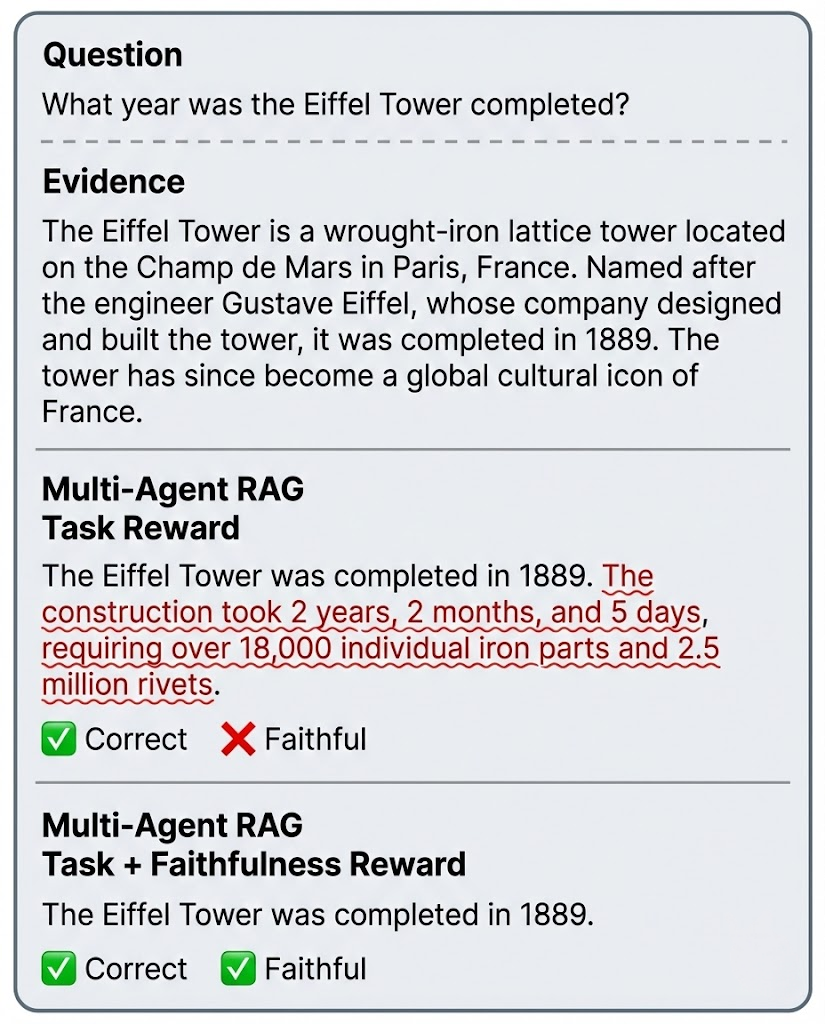}
\caption{A task-reward-only model can answer the main question correctly while still adding details not supported by the evidence (wavy underline). This gap motivates a reward signal that targets faithfulness directly.}
\label{fig:motivation}
\end{figure}

This mismatch is illustrated in Figure~\ref{fig:motivation}: an answer can score well on task metrics while adding details that the evidence does not mention. Existing supervision signals for faithfulness mostly operate on the final surface form. Reference-based metrics compare the answer to a gold reference, and LLM-as-a-judge approaches evaluate the generated answer post-hoc. Both evaluate the final answer rather than the generation process, and both are difficult to use as a dense reward during reinforcement learning: reference-based metrics only score how much the answer matches a gold span, while LLM-as-a-judge requires an external call for every rollout. A signal computed from the generator's own behavior during decoding would be cheaper to query and would capture the generation process itself, rather than only evaluating the final answer after it has been produced.

Recent interpretability work, however, suggests that the model already carries information about truthfulness in its internal representations~\citep{burns2023discovering,li2024iti,ho2025novo}. In particular, simple statistics over attention-head outputs can separate truthful from false continuations, which makes them a practical source of signal during decoding. Existing methods use this kind of internal signal in two ways: to detect hallucinations after the answer has been produced~\citep{su2024mind,chen2024inside}, or to steer activations during inference~\citep{li2024iti}. Both leave the policy unchanged and act only on the generator. In a multi-agent RAG pipeline, however, unsupported claims can also come from how the query is planned and which evidence is selected, so a faithfulness signal that only acts on the generator is not enough. This motivates our central question:

\begin{center}
\textbf{\textit{Can internal attention signals be turned into a reward that guides multi-agent RAG toward more faithful generation?}}
\end{center}

We answer this question with a hallucination-aware multi-agent RAG framework. The system decomposes RAG into three cooperative agents, a Query Planner, an Evidence Selector, and an Answer Generator, each implemented as a role-specific LoRA adapter~\citep{hu2022lora} over a shared, frozen backbone. A lightweight critic over attention-head norm features is trained once on TruthfulQA~\citep{lin2022truthfulqa} truthful/false pairs and then kept frozen, providing a cheap internal faithfulness signal that can be queried at every rollout. During cooperative PPO training, the critic's risk score is combined with the task reward to form a shared training signal, so that all three agents are trained against both answer correctness and an internal risk estimate, and the faithfulness signal reaches the query planner and evidence selector as well as the generator.

Our main contributions are as follows:
\begin{itemize}
\item A hallucination-aware multi-agent RAG framework that jointly optimizes query planning, evidence selection, and answer generation under a shared reward combining task correctness with an internal faithfulness signal.
\item A frozen risk critic over attention-head norm statistics, calibrated once on TruthfulQA and transferred across RAG benchmarks without retraining.
\item Empirical evidence across three backbones and three QA benchmarks showing consistent improvements on task metrics and reductions in hallucination rate over the strongest multi-agent baseline.
\end{itemize}

\section{Learning Setup}
\label{sec:method}

We present a hallucination-aware multi-agent RAG framework that uses
internal attention behavior as a reward signal during reinforcement
learning. The method is built around three design choices. First, we
decompose RAG into query planning, evidence selection, and answer
generation, so that each stage can be optimized as an explicit decision.
Second, we keep the backbone language model frozen and learn
role-specific LoRA adapters, which enables specialization while
preserving the internal features used by the critic. Third, we train an
offline attention-based critic and use its risk estimate as a
reward-shaping term that complements the task reward.

Due to space constraints, the related work is discussed in Appendix~\ref{sec:Related}.




\subsection{Problem Formulation}
\label{sec:probsetup}

We study knowledge-intensive question answering in a retrieval-augmented
setting. Each example contains a question $q$ and an evidence pool
$D=\{d_1,\ldots,d_K\}$, where $D$ may be returned by a retriever or
provided by the benchmark. The system outputs a natural-language answer
$a$ conditioned on the available evidence.

The learning objective has two parts: the answer should be correct with
respect to a reference, and it should be faithful to the provided
evidence. These goals are related but not identical---an answer can
overlap with the reference while still containing unsupported
claims~\cite{niu2024ragtruth,ji2023hallucination_survey}---and a
terminal task reward alone does not explicitly distinguish them.

Let $\pi_\Theta$ denote the frozen backbone language model, parameterized
by $\Theta$. Each agent $u \in \{Q,E,G\}$ is instantiated with a fixed
role prompt $p_u$ and a role-specific LoRA adapter $\Delta\Theta_u$,
which together define the agent policy
\begin{equation}
\pi_u(\cdot \mid x)
=
\pi_{\Theta+\Delta\Theta_u}(\cdot \mid p_u \oplus x),
\end{equation}
where $\oplus$ denotes text concatenation. The prompts $p_Q$, $p_E$,
and $p_G$ specify the instructions for query planning, evidence
selection, and answer generation, respectively.

\subsection{Framework Overview}
\label{sec:overview}

\begin{figure*}[t]
    \centering
    \includegraphics[width=\textwidth]{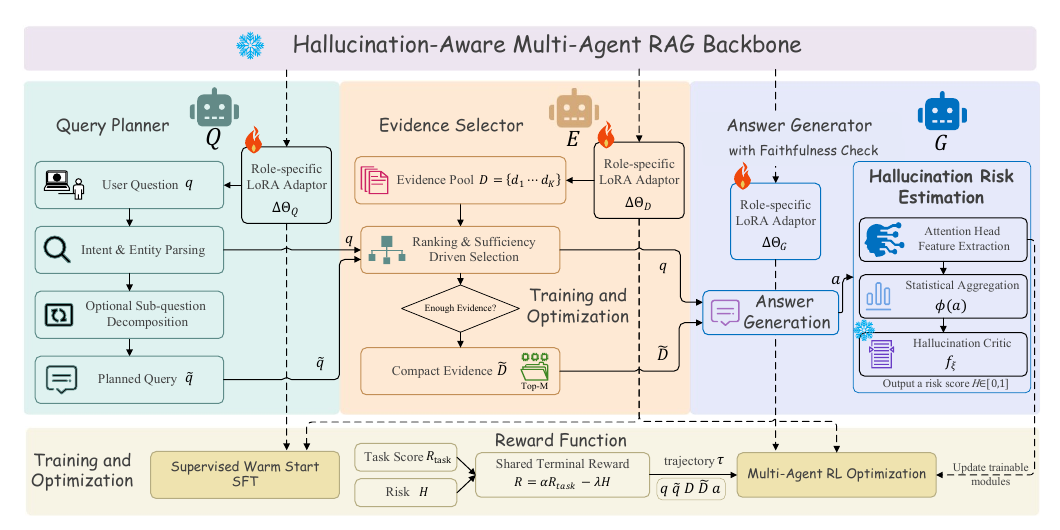}
    \caption{Overview of our hallucination-aware multi-agent RAG
    framework. The pipeline decomposes RAG into a query planner ($Q$),
    evidence selector ($E$), and answer generator ($G$), implemented as
    role-specific LoRA adapters on a frozen backbone. A frozen
    attention-based critic estimates a hallucination risk score from
    generator-side answer features; this score is combined with the
    task reward to form a shared terminal reward for cooperative PPO
    training.}
    \label{fig:framework}
\end{figure*}

Figure~\ref{fig:framework} illustrates the overall framework. We
formulate RAG as a turn-based cooperative decision process with three
agents executed in a fixed order: Agent Q rewrites or plans the query,
Agent E selects evidence from the candidate pool, and Agent G generates
the final answer. The episode receives a single terminal reward shared
by all agents.

The joint trajectory of one episode is
$\tau=(q,\tilde{q},D,\tilde{D},a)$, where $\tilde{q}$ is the planned
query, $\tilde{D}\subseteq D$ is the selected evidence subset, and $a$
is the final answer.

Following the notation in Section~\ref{sec:probsetup}, all agents are
instantiated from the same frozen backbone $\pi_\Theta$ and differ only
through their role-specific prompts and LoRA adapters. This design keeps
the framework parameter-efficient while allowing each agent to
specialize for its own role. Keeping the backbone and critic frozen
during RL also preserves the feature space on which the critic was
calibrated.

At the end of an episode, the frozen critic outputs a hallucination risk
score $H\in[0,1]$ from the generator-side answer representation. The
shared terminal reward is
\begin{equation}
R = \alpha R_{\text{task}} - \lambda H,
\label{eq:overview_reward}
\end{equation}
where $\alpha$ and $\lambda$ control the relative strength of task
performance and hallucination-risk penalization.

\subsection{Multi-Agent RAG Policies}
\label{sec:agents}

The three agents share the same language-model interface but receive
different prompts, observations, and LoRA adapters. This preserves a
unified text-generation interface while assigning each policy a
distinct functional role.


\paragraph{Agent Q: Query Planner.}
Agent Q converts the original question into a clarified query or
lightweight plan, $\tilde{q}\sim\pi_Q(\cdot\mid q)$. The planned query
makes salient entities, constraints, or sub-questions explicit. In our
fixed-evidence setting, Agent Q does not trigger a new retrieval call,
but provides a clearer conditioning signal for evidence selection and
answer generation.

\paragraph{Agent E: Evidence Selector.}
Agent E chooses which evidence should condition the generator. This
stage matters because irrelevant or redundant passages distract the
generator, while missing key evidence forces it to rely on parametric
memory.

Given $q$, $\tilde{q}$, and $D$, Agent E emits a textual selection of
document identifiers, which is parsed into an evidence subset:
\begin{equation}
\tilde{D} \sim \pi_E(\cdot \mid q, \tilde{q}, D),
\quad
\tilde{D} = \{d_{i_1}, \ldots, d_{i_M}\}.
\end{equation}
We use a canonical textual schema for the selection action; invalid or
duplicate identifiers are handled by the parser. This keeps evidence
selection within the same language-model action interface as the other
two agents.

\paragraph{Agent G: Answer Generator.}
Agent G is the only module that produces the free-form answer, so we define its policy over the question and selected evidence, and compute the hallucination-risk signal from its answer-side internal features: $a\sim\pi_G(\cdot\mid q,\tilde{D})$. Although Agent Q and Agent E do not generate the
answer directly, they are trained under the same shared reward, so their intermediate decisions are shaped toward inputs that let the generator produce correct and faithful answers.

\subsection{Attention-Based Hallucination Critic}
\label{sec:critic}

The hallucination critic maps internal attention-head features to a
scalar risk score. It is trained offline on a truthfulness calibration
set and held frozen during policy optimization, providing a stable
signal that guides the policy toward answer patterns associated with
truthful continuations.

\paragraph{Attention Heads as Truth Indicators.}
Prior work has shown that transformer internals carry information
correlated with
truthfulness~\cite{burns2023discovering,li2024iti,ho2025novo}. In
particular, NoVo~\cite{ho2025novo} demonstrates that the norms of
certain attention-head outputs can distinguish truthful from false
answers on TruthfulQA. Building on this observation, we use attention
head norms as compact internal features. Let
$\mathbf{c}^{l,h}_{t}$ denote the output vector of head $h$ in layer
$l$ at answer-token position $t$. We define the token-level head-norm
feature as
\begin{equation}
x_t^{l,h} = \|\mathbf{c}^{l,h}_{t}\|_2 .
\end{equation}
This statistic is simple, model-internal, and available without an
external hallucination annotator at RL time.

\paragraph{Head Feature Extraction.}
For each generated answer, we run a generator-side scoring pass and
collect attention-head outputs over the answer tokens. Let
$a=(a_1,\ldots,a_T)$ denote the generated answer token sequence. For
each selected head, we aggregate token-level head norms over the answer
by mean pooling:
\begin{equation}
\bar{x}^{l,h}(a)
=
\frac{1}{T}\sum_{t=1}^{T} x_t^{l,h}.
\end{equation}
The fixed-dimensional critic input is then
\begin{equation}
\phi(a)
=
\left[
\bar{x}^{l,h}(a)
\right]_{(l,h)\in \mathcal{H}_{\mathrm{sel}}},
\end{equation}
where $\mathcal{H}_{\mathrm{sel}}$ is the selected head set. Thus,
$\phi(a)$ contains one pooled norm feature for each selected head.
Additional implementation details are provided in
Appendix~\ref{sec:implementation_details}.

\paragraph{Truth-Correlated Head Selection.}
Not all heads carry useful truthfulness-related signal, so we select
heads using a TruthfulQA-derived calibration set~\cite{lin2022truthfulqa}.
Each calibration example contains a truthful answer and a false answer
for the same question. For each head, we measure whether its pooled norm
feature separates the two answers in a consistent direction.

For each truthful/false pair $(s_i^+,s_i^-)$, we compute
\begin{equation}
\Delta \bar{x}_i^{l,h}
=
\bar{x}^{l,h}(s_i^+) - \bar{x}^{l,h}(s_i^-).
\end{equation}
We then define the pair-level vote as
\begin{equation}
v_i^{l,h}
=
\begin{cases}
1, & \mathrm{sign}^{l,h} \cdot \Delta \bar{x}_i^{l,h} > 0, \\
0, & \text{otherwise},
\end{cases}
\end{equation}
where $\mathrm{sign}^{l,h}\in\{+1,-1\}$ indicates the direction in which
the head is correlated with truthful answers. The discrimination
accuracy of each head is computed as
\begin{equation}
\mathrm{Acc}^{l,h}
=
\frac{1}{N}\sum_{i=1}^{N} v_i^{l,h}.
\end{equation}
Heads exceeding an accuracy threshold are retained in
$\mathcal{H}_{\mathrm{sel}}$, yielding a compact set of
truth-correlated heads as the critic's input features.

\paragraph{Critic Architecture and Training.}
The critic is a lightweight MLP that maps the aggregated head features
to a hallucination-risk probability:
\begin{equation}
H = f_\xi(\phi(a)) \in [0,1].
\end{equation}
We train $f_\xi$ with binary cross-entropy on the calibration pairs,
assigning higher risk to false answers and lower risk to truthful
answers. Once trained, the critic is fixed and used to provide the
scalar risk term $H$ during RL. Freezing both the critic and the
backbone avoids a moving reward function during policy optimization and
keeps the head-feature distribution stable across training.

\subsection{Hallucination-Aware Multi-Agent RL}
\label{sec:rl}

We integrate the critic into RL through a shared terminal reward. This
keeps credit assignment simple: every agent receives the same scalar
feedback, but each updates only its own role-specific adapter.

\paragraph{Supervised warm start.}
Before RL, we warm-start the three adapters with supervised
instruction-following data, teaching the model to produce valid query
plans, evidence selections, and answers under the corresponding
prompts. RL then refines these role policies under the
hallucination-aware reward. This separation keeps the exploration
    problem tractable and avoids relying on RL to discover the basic
action formats from scratch.


\paragraph{Reward.}
After Agent G produces answer $a$, we compute the task reward as
\begin{equation}
R_{\text{task}} = \mathrm{F1}(a, a^*),
\end{equation}
where $a^*$ is the reference answer. The frozen critic then estimates
the hallucination risk $H=f_\xi(\phi(a))$ from the answer's head-norm
features. The final shared terminal reward follows
Eq.~\ref{eq:overview_reward}, where the task reward is combined with the
critic-based hallucination penalty. The coefficients $\alpha$ and
$\lambda$ are tuned on a development split.

\paragraph{PPO-Style Cooperative Multi-Agent Optimization.}
We optimize the three agents with a cooperative PPO objective. All
agents receive the same terminal reward defined in
Eq.~\ref{eq:overview_reward}. We compute a shared advantage
$A=R-b$, where $b$ is an exponential-moving-average baseline of recent
episode rewards. The advantage is batch-normalized when its variance is
sufficiently large.

Each agent $i \in \{Q,E,G\}$ emits a textual action
$z_i=(y_{i,1},\ldots,y_{i,T_i})$ under its role-specific observation
$o_i$. Let $\theta_i=\Theta+\Delta\Theta_i$ denote the frozen backbone
combined with the LoRA adapter of agent $i$. During the PPO update, we
recompute the length-normalized log-likelihood of the sampled action:
\begin{equation}
\bar{\ell}_{\theta_i}(z_i \mid o_i)
=
\frac{1}{T_i}
\sum_{t=1}^{T_i}
\log \pi_{\theta_i}
\left(
y_{i,t} \mid o_i, y_{i,<t}
\right).
\end{equation}
Here, $y_{i,t}$ is the $t$-th token of agent $i$'s textual action, and
$\pi_{\theta_i}$ denotes the token-level distribution induced by that
agent's policy.

The sequence-level likelihood ratio is
\begin{equation}
r_i =
\exp\!\left(
\bar{\ell}_{\theta_i}(z_i \mid o_i)
-
\bar{\ell}_{\theta_i^{\mathrm{old}}}(z_i \mid o_i)
\right),
\end{equation}
where $\theta_i^{\mathrm{old}}$ denotes the rollout policy used to
generate $z_i$. The clipped PPO loss for each agent is
\begin{equation}
\mathcal{L}_{\mathrm{PPO}}^{(i)}
=
-\mathbb{E}\!\left[
\min\!\left(
r_i A,\;
\mathrm{clip}(r_i, 1-\epsilon, 1+\epsilon) A
\right)
\right].
\end{equation}
We also add a small sequence-level regularizer to discourage abrupt
deviation from the rollout policy:
\begin{equation}
\mathcal{L}_{\mathrm{reg}}^{(i)}
=
\left(
\bar{\ell}_{\theta_i}(z_i \mid o_i)
-
\bar{\ell}_{\theta_i^{\mathrm{old}}}(z_i \mid o_i)
\right)^2.
\end{equation}
The total objective sums over the three agents:
\begin{equation}
\mathcal{L}
=
\sum_{i \in \{Q,E,G\}}
\left(
\mathcal{L}_{\mathrm{PPO}}^{(i)}
+
\rho \mathcal{L}_{\mathrm{reg}}^{(i)}
\right).
\end{equation}
We update only the role-specific LoRA adapters
$\{\Delta\Theta_Q,\Delta\Theta_E,\Delta\Theta_G\}$ with separate
optimizers, while keeping both the backbone $\Theta$ and the critic
$f_\xi$ frozen.

This cooperative objective allows all agents to learn from the same
final outcome: the query planner is rewarded when its plan improves
downstream evidence use, the evidence selector when its selected
evidence supports grounded answering, and the generator when its final
answer is both correct and low-risk under the frozen critic.

\section{Experiments}
\label{sec:experiments}

\subsection{Experimental Setup}
\label{sec:setup}

We evaluate our framework on three knowledge-intensive QA benchmarks:
RAGTruth~\citep{niu2024ragtruth}, HotpotQA~\citep{yang2018hotpotqa},
and 2WikiMultiHopQA~\citep{ho2020constructing}. All methods are tested
under a fixed-evidence setting, where the same evidence pool is provided
to each system, so that the comparison focuses on policy learning rather
than retriever variability. We instantiate our method with three
backbones: Qwen3-8B, Llama3-8B, and Llama2-13B. The baselines include
vanilla RAG variants, SelfRAG~\citep{asai2024selfrag}, and recent
multi-agent RAG methods, including MAIN-RAG~\citep{chang-etal-2025-mainrag},
RAG-DDR~\citep{li2024ragddr}, MA-RAG~\citep{nguyen2025marag}, and
MMOA-RAG~\citep{chen2026mmoa}. We report Recall/EM/F1 on RAGTruth and
Acc/EM/F1 on HotpotQA and 2WikiMultiHopQA, using R/A to denote Recall
for RAGTruth and Accuracy for the two multi-hop QA datasets in averaged
columns. We additionally report Answer Hallucination Rate (AHR) on
RAGTruth to measure the fraction of generated answers containing claims
unsupported by the provided evidence. Further details on datasets,
implementation settings, baselines, and evaluation protocols are
provided in Appendix~\ref{Experimental_Details}.

\subsection{Main Results}
\label{sec:main_results}

\begin{table*}[tb]
  \centering
  \small
  \setlength{\tabcolsep}{4pt}

  \begin{tabular}{l ccc ccc ccc ccc}
    \toprule
    & \multicolumn{3}{c}{\textbf{RAGTruth}}
    & \multicolumn{3}{c}{\textbf{HotpotQA}}
    & \multicolumn{3}{c}{\textbf{2WikiMultiHopQA}}
    & \multicolumn{3}{c}{\textbf{Average}} \\
    \cmidrule(lr){2-4}\cmidrule(lr){5-7}\cmidrule(lr){8-10}\cmidrule(lr){11-13}
    \textbf{Method}
    & Recall & EM & F1
    & Acc & EM & F1
    & Acc & EM & F1
    & R/A & EM & F1 \\
    \midrule

    \multicolumn{13}{c}{\textit{Qwen3-8B}} \\
    \midrule
    Zeroshot w/ SFT      & 29.08 & 25.31 & 33.91 & 28.78 & 24.68 & 33.47 & 30.21 & 23.96 & 36.42 & 29.36 & 24.65 & 34.60 \\
    Vanilla RAG          & 31.99 & 24.62 & 34.67 & 34.94 & 16.91 & 38.84 & 34.09 & 25.42 & 36.56 & 33.67 & 22.32 & 36.69 \\
    Vanilla RAG w/ SFT   & 37.18 & 33.31 & 44.49 & 38.47 & 35.28 & 43.36 & 38.41 & 33.74 & 44.45 & 38.02 & 34.11 & 44.10 \\
    SelfRAG              & 34.42 & 31.77 & 42.93 & 37.32 & 36.39 & 38.68 & 32.35 & 28.73 & 43.04 & 34.70 & 32.30 & 41.55 \\
    MAIN-RAG             & 38.69 & 36.62 & 45.49 & 38.02 & 35.73 & 43.51 & 39.13 & 35.37 & 44.78 & 38.61 & 35.91 & 44.59 \\
    MA-RAG               & 41.02 & 37.24 & 43.31 & 40.58 & 37.92 & 45.12 & 40.15 & 36.84 & 45.83 & 40.58 & 37.33 & 44.75 \\
    RAG-DDR              & 47.20 & 36.65 & 46.26 & 46.49 & 38.45 & 47.18 & 47.25 & 42.55 & 49.83 & 46.98 & 39.22 & 47.76 \\
    MMOA-RAG             & 46.01 & \textbf{39.56} & 46.30 & 47.12 & 37.68 & 46.74 & 47.86 & 43.95 & 50.05 & 47.00 & 40.40 & 47.70 \\
    \rowcolor{myteal!30}
    \textbf{Ours}        & \textbf{48.08} & 38.18 & \textbf{47.89} & \textbf{49.34} & \textbf{40.02} & \textbf{48.44} & \textbf{50.62} & \textbf{45.84} & \textbf{50.63} & \textbf{49.35} & \textbf{41.35} & \textbf{48.99} \\
    \midrule

    \multicolumn{13}{c}{\textit{Llama3-8B}} \\
    \midrule
    Zeroshot w/ SFT      & 28.45 & 24.78 & 34.52 & 27.92 & 23.85 & 32.68 & 29.38 & 23.15 & 35.59 & 28.58 & 23.93 & 34.26 \\
    Vanilla RAG          & 31.24 & 23.89 & 33.94 & 34.11 & 16.08 & 25.01 & 33.26 & 24.59 & 35.73 & 32.87 & 21.52 & 31.56 \\
    Vanilla RAG w/ SFT   & 39.52 & 35.67 & 43.86 & 41.64 & 34.45 & 46.53 & 37.58 & 32.91 & 47.53 & 39.58 & 34.34 & 45.97 \\
    SelfRAG              & 33.79 & 31.14 & 40.37 & 38.49 & 32.56 & 42.03 & 31.52 & 27.87 & 42.21 & 34.60 & 30.52 & 41.54 \\
    MAIN-RAG             & 39.96 & 37.87 & 42.86 & 43.19 & 33.94 & 44.68 & 38.34 & 34.54 & 47.95 & 40.50 & 35.45 & 45.16 \\
    MA-RAG               & 40.32 & 35.93 & 43.91 & 48.78 & 35.78 & 46.46 & 49.18 & 44.76 & 51.13 & 46.09 & 38.82 & 47.17 \\
    RAG-DDR              & 46.42 & 39.42 & 47.66 & 49.95 & 36.69 & 48.57 & 49.02 & 44.92 & 51.76 & 48.46 & 40.34 & 49.33 \\
    MMOA-RAG             & 47.05 & 39.18 & 46.94 & 49.83 & 38.91 & 49.81 & 49.56 & 45.28 & \textbf{52.06} & 48.81 & 41.12 & 49.60 \\
    \rowcolor{myteal!30}
    \textbf{Ours}        & \textbf{47.74} & \textbf{40.63} & \textbf{48.38} & \textbf{50.54} & \textbf{41.37} & \textbf{49.95} & \textbf{50.57} & \textbf{45.69} & 51.98 & \textbf{49.62} & \textbf{42.56} & \textbf{50.10} \\
    \midrule

    \multicolumn{13}{c}{\textit{Llama2-13B}} \\
    \midrule
    Zeroshot w/ SFT      & 28.71 & 20.35 & 28.28 & 28.71 & 22.72 & 29.51 & 27.17 & 23.33 & 28.46 & 28.20 & 22.13 & 28.75 \\
    Vanilla RAG          & 30.48 & 23.66 & 32.60 & 31.98 & 16.95 & 30.32 & 31.11 & 16.67 & 24.17 & 31.19 & 19.09 & 29.03 \\
    Vanilla RAG w/ SFT   & 41.22 & 32.34 & 41.87 & 38.83 & 32.84 & 40.46 & 39.45 & 34.78 & 41.41 & 39.83 & 33.32 & 41.25 \\
    SelfRAG              & 30.46 & 27.81 & 33.97 & 30.36 & 29.43 & 33.98 & 32.39 & 29.74 & 34.08 & 31.07 & 28.99 & 34.01 \\
    MAIN-RAG             & 41.73 & 34.64 & 41.53 & 40.06 & 37.77 & 40.55 & 40.17 & 36.41 & 40.82 & 40.65 & 36.27 & 40.97 \\
    MA-RAG               & 41.09 & 37.84 & 43.58 & 42.65 & 38.65 & 43.33 & 41.05 & 36.57 & 43.80 & 41.60 & 37.69 & 43.57 \\
    RAG-DDR              & 42.19 & 38.19 & 44.33 & 42.77 & 38.56 & 44.44 & 41.89 & 37.79 & 44.63 & 42.28 & 38.18 & 44.47 \\
    MMOA-RAG             & 42.68 & 38.73 & 44.82 & 43.12 & 39.08 & \textbf{44.97} & 42.35 & 38.19 & 45.13 & 42.72 & 38.67 & 44.97 \\
    \rowcolor{myteal!30}
    \textbf{Ours}        & \textbf{46.09} & \textbf{42.73} & \textbf{46.92} & \textbf{45.55} & \textbf{40.47} & 44.54 & \textbf{48.79} & \textbf{40.50} & \textbf{46.99} & \textbf{46.81} & \textbf{41.23} & \textbf{46.15} \\
    \bottomrule
  \end{tabular}
  \caption{Main results across three datasets and backbones. The first
  metric is Recall for RAGTruth and Accuracy for HotpotQA and
  2WikiMultiHopQA; rightmost columns average the first metric (R/A),
  EM, and F1 across datasets. Best results in \textbf{bold}.}
  \label{tab:main_results}
\end{table*}

Table~\ref{tab:main_results} reports the main comparison. Our method
achieves the strongest average performance for all three backbones,
with the largest gains on the first metric (RAGTruth Recall or QA
Accuracy) and EM. On Qwen3-8B, it improves over the strongest
multi-agent baseline, MMOA-RAG, by +2.35 R/A, +0.95 EM, and +1.29 F1 on
average. The gains are visible across all three datasets: +2.07 Recall on RAGTruth, +2.22 Accuracy and +2.34 EM on HotpotQA, and
+2.76 Accuracy and +1.89 EM on 2WikiMultiHopQA.

On Llama3-8B and Llama2-13B, our method also leads on the average across
all three metrics, with consistent gains on Recall/Accuracy and EM.
These results indicate that the improvements from the hallucination-aware
reward generalize across backbone architectures and pre-training data,
rather than depending on a particular model family.

Compared with other multi-agent and reward-based methods, the
difference lies in the reward signal rather than in pipeline structure. The internal hallucination-risk estimate provides supervision that
surface-form task rewards do not capture, and the average gains in EM
and F1, which are more sensitive to deviations from the reference are consistent with this interpretation.

\subsection{Faithfulness Analysis}
\label{sec:faithfulness}

Table~\ref{tab:ahr} reports Answer Hallucination Rate (AHR) on
RAGTruth, which directly measures the fraction of responses containing
unsupported claims. Our method achieves the lowest AHR across all three
backbones, showing a consistent reduction in unsupported generation
compared with the baselines. The improvement is most pronounced on
Qwen3-8B, where the absolute reduction over MMOA-RAG reaches 10.74
points. The same pattern also holds on Llama3-8B and Llama2-13B, with
absolute reductions of 5.62 and 3.34 points, respectively. These results
indicate that the proposed framework improves not only
task-level evaluation scores, but also the practical
faithfulness of generated answers.

Overall, the AHR results indicate that the hallucination-aware critic
provides a useful reward-shaping signal for improving faithfulness. The
internal risk estimate encourages the model to move away from
unsupported generations, while the improvements in
Table~\ref{tab:main_results} show that this does not come at the expense
of task performance. Instead, faithfulness and task-level effectiveness
improve in parallel, suggesting that the proposed reward design aligns
well with the overall objective of grounded answer generation.


\begin{table}[t]
    \centering
    \small
    \setlength{\tabcolsep}{3pt}
    \resizebox{\columnwidth}{!}{
    \begin{tabular}{lccc}
        \toprule
        Method & Qwen3-8B & Llama3-8B & Llama2-13B \\
        \midrule
        Zeroshot w/ SFT      & 76.35 & 74.62 & 73.81 \\
        Vanilla RAG          & 51.29 & 50.36 & 44.27 \\
        Vanilla RAG w/ SFT   & 53.06 & 51.62 & 52.37 \\
        SelfRAG              & 34.42 & 37.29 & 36.15 \\
        MAIN-RAG             & 40.31 & 42.11 & 41.02 \\
        RAG-DDR              & 32.46 & 34.02 & 35.68 \\
        MA-RAG               & 32.99 & 35.32 & 37.59 \\
        MMOA-RAG             & 31.98 & 30.27 & 31.03 \\
        \rowcolor{myteal!30}
        \textbf{Ours}        & \textbf{21.24} & \textbf{24.65} & \textbf{27.69} \\
        \bottomrule
    \end{tabular}
    }
    \caption{Answer Hallucination Rate (AHR, \%; lower is better) on RAGTruth.}
    \label{tab:ahr}
\end{table}

\subsection{Ablation Studies}
\label{sec:ablation}

\begin{figure*}[tb]
    \centering
    \includegraphics[width=\textwidth]{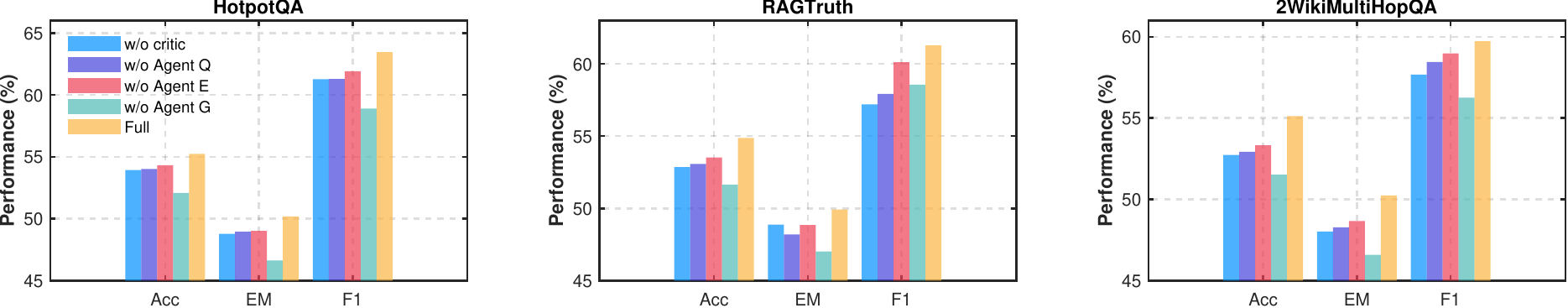}
    \caption{Ablation study of our framework
    (Qwen3-8B).}
    \label{fig:ablation}
\end{figure*}

Table~\ref{tab:ablation} and Figure~\ref{fig:ablation} analyze the
contributions of the hallucination critic and the role-specific agent
updates on Qwen3-8B. For the agent-update ablations, the full
query--evidence--generation pipeline is kept unchanged at inference
time. The ablated agent still participates in the pipeline, but its
LoRA adapter is frozen at the SFT initialization during PPO, while the
other agents continue to update. This allows us to study the effect of
RL on each role without changing the overall system interface.

Removing the hallucination critic leads to a clear performance drop:
average R/A decreases by 3.31 points and average F1 decreases by 2.59
points. This directly tests whether the TruthfulQA-calibrated internal
risk signal contributes information beyond the task reward alone, and
the observed drop suggests that it provides useful guidance during
policy optimization. The agent-update ablations show a similar pattern:
freezing any role degrades performance, with the generator update having
the largest effect. This is consistent with the generator's direct role
in producing the final answer, while the smaller but consistent drops
from freezing the query planner or evidence selector indicate that
upstream decisions also contribute to the final generation quality.

\subsection{Critic Diagnostics and Head Analysis}
\label{sec:critic_analysis}

\paragraph{Calibration source and supporting evidence.}
We calibrate the critic on TruthfulQA rather than on dataset-specific
faithfulness annotations. TruthfulQA provides paired truthful and false
answers outside the RAGTruth retrieval distribution, which allows the
critic to be used as a transferable internal risk signal instead of a
dataset-specific hallucination classifier. This choice also avoids
training the reward model directly on the annotations used for
faithfulness evaluation.

We evaluate the critic from two perspectives. On held-out TruthfulQA
pairs, the best calibrated critic achieves 75.92\% AUROC, indicating
that the selected attention-head norm features contain a non-random
truthfulness signal. When used for RAG policy optimization, the critic
also produces risk scores in an actionable range, rather than
saturating near zero or one. Together with the transfer diagnostics in
Appendix~\ref{sec:critic_transfer_appendix}, these results support our
use of the critic as a lightweight auxiliary reward signal, rather than
as a standalone hallucination detector.

We further examine which heads contribute to this signal and how their
norms behave over generated answers.

\begin{figure}[t]
    \centering
    \includegraphics[width=\columnwidth]{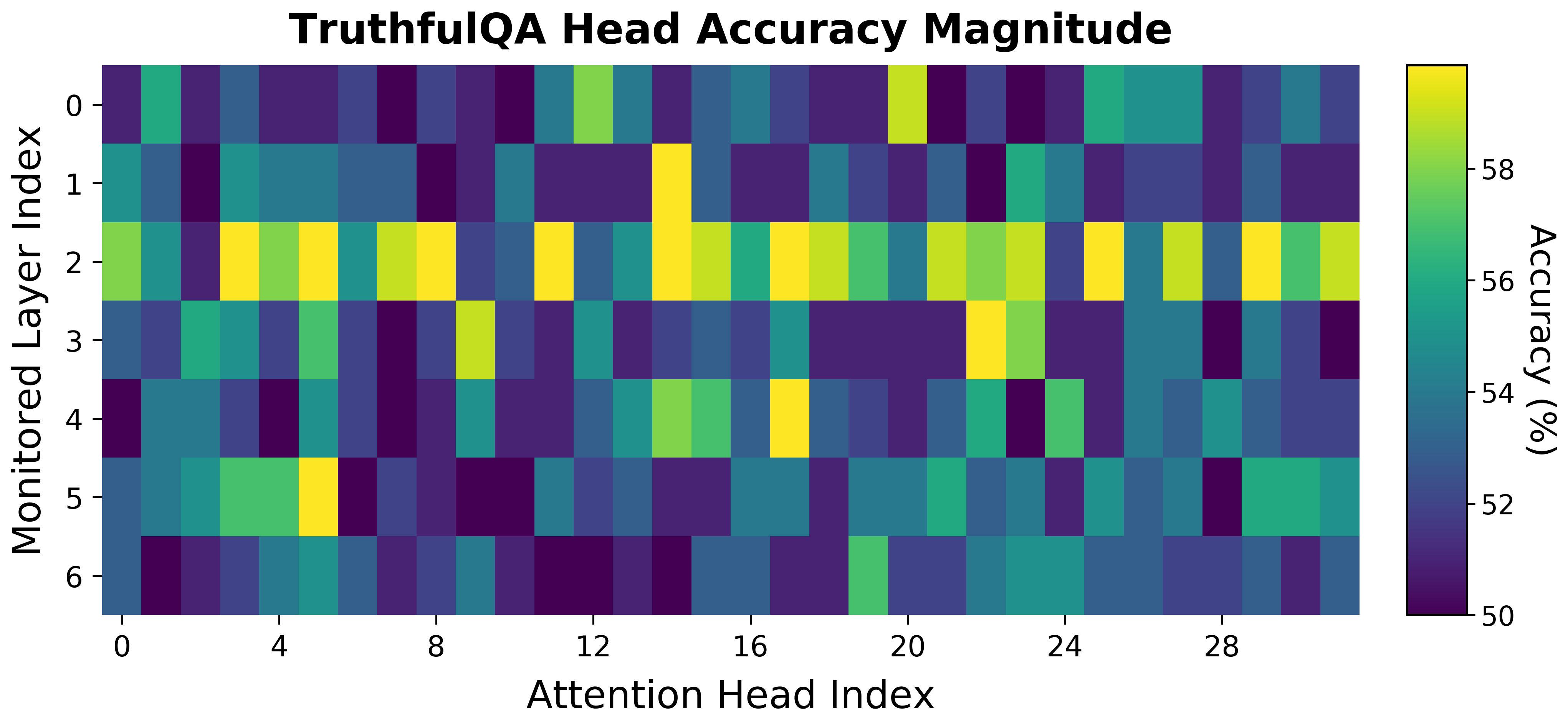}
    \caption{Head discrimination map on TruthfulQA. Each cell encodes
    the stronger of two head-level vote accuracies. Discriminative
    behavior is concentrated in a subset of heads spanning multiple
    layers.}
    \label{fig:head_acc}
\end{figure}

\paragraph{Head discrimination map.}
Figure~\ref{fig:head_acc} visualizes the discrimination strength of
individual attention heads on paired truthful and false answers. The
signal is concentrated in a subset of heads across multiple layers,
rather than being uniformly distributed throughout the model. This
pattern supports the head-selection procedure in
Section~\ref{sec:critic}, which retains the heads with the strongest
truth-correlated behavior as critic inputs.

\begin{figure}[t]
    \centering
    \includegraphics[width=\columnwidth]{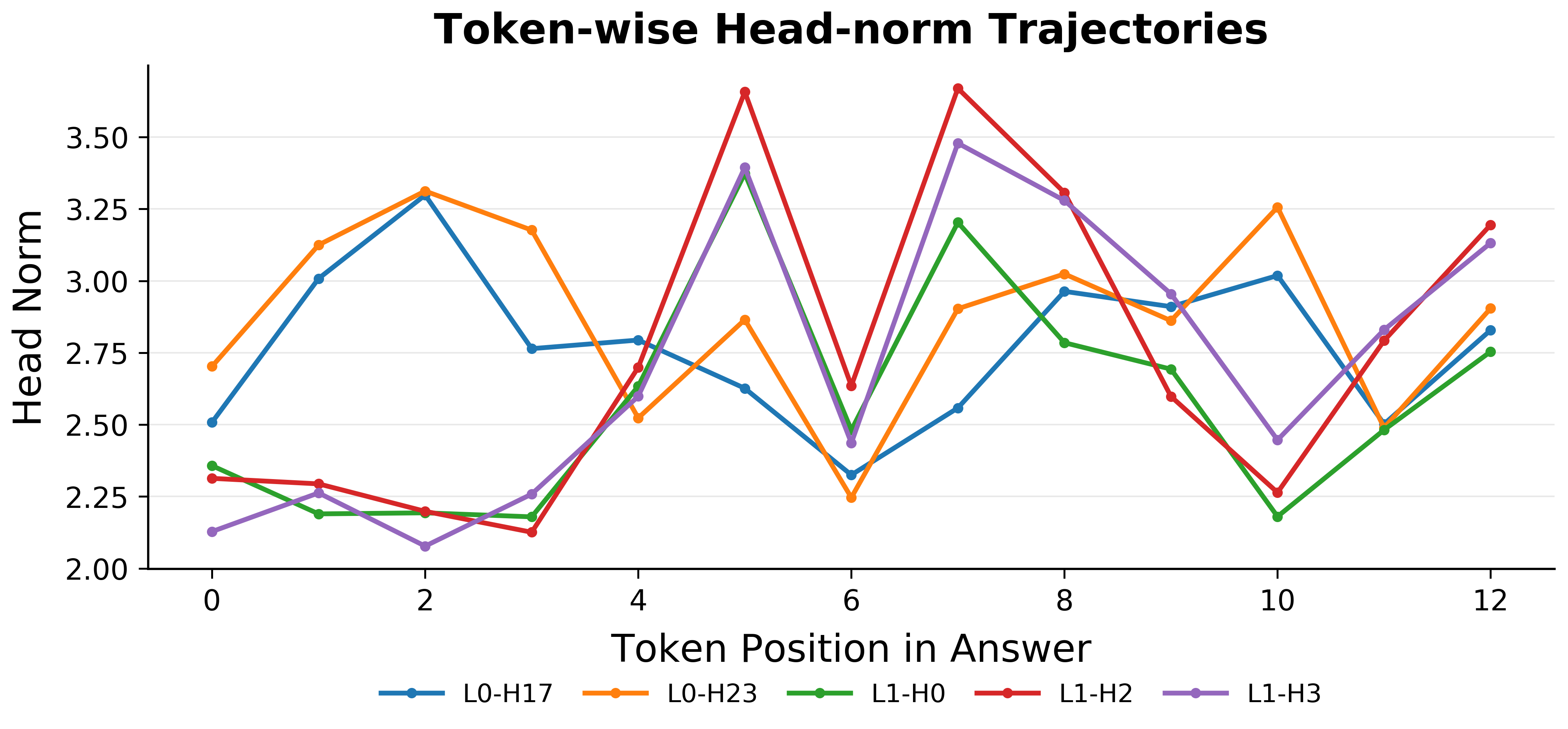}
    \caption{Token-wise head-norm trajectories for representative
    heads. Local changes appear near answer-salient tokens, indicating
    that head norms respond to content-bearing positions.}
    \label{fig:head_norm_curves}
\end{figure}

\paragraph{Token-wise dynamics.}
Figure~\ref{fig:head_norm_curves} shows token-wise head-norm
trajectories for representative selected heads. The curves exhibit
local changes around answer-salient tokens, such as entity mentions and
factual predicates. This suggests that the selected head norms are
sensitive to content-bearing positions in the answer, making them
reasonable internal features for faithfulness-oriented reward shaping.

\section{Conclusion}
\label{sec:conclusion}

We presented a hallucination-aware multi-agent RAG framework that uses
internal attention-head diagnostics as an auxiliary reward signal for
reinforcement learning. The framework decomposes RAG into query
planning, evidence selection, and answer generation, and trains
role-specific LoRA adapters with a shared reward that combines task
performance with a frozen critic's hallucination-risk estimate.
Experiments on RAGTruth, HotpotQA, and 2WikiMultiHopQA show consistent
improvements across Qwen3-8B, Llama3-8B, and Llama2-13B, while also
reducing answer hallucination on RAGTruth. These results suggest that
lightweight model-internal signals can effectively shape RAG policies
toward more faithful generation when used as a complementary reward
rather than as a replacement for task supervision. More broadly, our
findings indicate that the internal behavior of the generator can
provide useful training-time feedback for grounded generation, beyond
what is captured by answer-overlap metrics alone. We hope this encourages
future work on combining external evidence supervision with
model-internal diagnostics for more reliable RAG systems.



\section*{Limitations}
\label{sec:limitations}

Although our framework shows that model-internal signals can provide
effective reward shaping for hallucination-aware multi-agent RAG, this
work focuses on a lightweight and stable instantiation of this idea.
First, the critic uses attention-head norm features as its internal
signal, which keeps the reward inexpensive to compute during RL and
allows the critic to remain simple and stable. Future work could examine
whether richer diagnostics, such as hidden-state dynamics, MLP
activations, or structured attention patterns, provide complementary
faithfulness-related information. Second, we keep the critic frozen during policy optimization. This
design preserves a fixed reward function and avoids changing the
calibration of the internal features while the agents are being updated.
Future work could explore adaptive critic-update strategies that track
policy changes while preserving training stability.

\section*{Acknowledgments}
Runsong Jia, Mengjia Wu, and Yi Zhang were supported by the Australian Commonwealth Scientific and Industrial Research Organization (CSIRO) in conjunction with the National Science Foundation (NSF) of the United States, under CSIRO-NSF \#2303037.
Mengjia Wu was also supported by Australian Research Council under Discovery Early Career Researcher Award DE260101493.


\bibliography{custom}

\appendix

\section{Appendix}
\label{sec:appendix}

\subsection{Related Work}
\label{sec:Related}

Our work draws on three lines of research: faithfulness in RAG systems,
reinforcement learning for cooperative LLM agents, and internal signals for
truthfulness. We situate our contribution by two questions: where the
supervision signal can come from (external text or internal
representations), and when it is applied (inference-time correction or
training-time reward shaping).

\subsubsection{Faithfulness and Multi-Agent Optimization in RAG}

A growing body of work targets hallucination in
RAG~\cite{lewis2020rag,gao2024rag_survey,ji2023hallucination_survey,niu2024ragtruth,jia2026evidence}.
Inference-time approaches augment the pipeline with self-critique or
correction: Self-RAG~\cite{asai2024selfrag} emits reflection tokens that
trigger adaptive retrieval and self-evaluation, while
CRAG~\cite{yan2024crag} introduces a lightweight evaluator that detects
insufficient retrieval and falls back on web search.
RefChecker~\cite{hu2024refchecker} contributes a fine-grained verification
scheme by decomposing responses into claim triplets. On the training side,
RAG-HAT~\cite{song-etal-2024-rag-hat} demonstrates that DPO over
faithful/unfaithful pairs can effectively reduce hallucinations, and
RAFT~\cite{zhang2024raft} shows how domain-specific RAG benefits from
targeted fine-tuning.

A parallel direction decomposes RAG into cooperating modules and optimizes
them jointly. Query rewriting~\cite{ma-etal-2023-query-rewriting,mao2024rafe},
iterative retrieval-reasoning~\cite{trivedi-etal-2023-ircot}, and adaptive
retrieval~\cite{jeong2024adaptive_rag} establish that individual stages
benefit from being treated as learnable. Multi-agent formulations extend
this idea: MMOA-RAG~\cite{chen2026mmoa} casts RAG as cooperative MARL with
shared F1-based rewards; RankRAG~\cite{yu2024rankrag} instruction-tunes a
single model to perform both ranking and generation; and
Chain-of-Note~\cite{yu2024chainofnote} bridges retrieval and generation
with explicit reading notes. Collectively, these methods demonstrate that
decomposing RAG into learnable stages produces substantial gains over
monolithic pipelines, and that text-level rewards such as F1 or preference
judgments effectively optimize answer overlap. Our work builds directly
on this line, adopting a similar multi-agent decomposition and PPO
training framework, and extends it along a complementary axis: we examine
whether a reward signal drawn from the model's own internal computations
can supplement these strong text-level objectives.

\subsubsection{Reinforcement Learning for LLM Alignment}

RL has become standard machinery for aligning LLMs.
InstructGPT~\cite{ouyang2022instructgpt} established the
SFT-then-PPO~\cite{schulman2017ppo} pipeline that underpins much of
subsequent work; DPO~\cite{rafailov2023dpo} offered a streamlined
alternative by directly optimizing over preference pairs. For reasoning,
GRPO~\cite{shao2024deepseekmath,deepseek2025r1} shows that strong
policies can be trained without a separate value network by estimating
baselines from group statistics. Constitutional AI~\cite{bai2022constitutional}
opens up an additional dimension by deriving training signal from
model-generated critique. In the RAG context, RAG-DDR~\cite{li2024ragddr}
contributes a differentiable data rewards approach for end-to-end training. And other works~\cite{xiong2026scaling,jia2026agent} show the ability of Agentic RAG and RL alignment.
This line of work has converged on text-domain rewards: preferences,
ratings, or match scores, which align language models with broad
behavioral targets effectively. We retain the PPO machinery these methods
established and explore a complementary source of supervision: signals
computed from the policy's own internal representations during generation.

\subsubsection{Internal Signals for Truthfulness}

A separate line of interpretability work has shown that transformer
hidden states carry meaningful information about output
truthfulness~\cite{burns2023discovering,li2024iti,ho2025novo,su2024mind,chen2024inside}.
Burns et al.~\cite{burns2023discovering} demonstrate that latent
knowledge can be extracted from activations without supervision;
ITI~\cite{li2024iti} reveals that shifting attention head activations at
inference time elicits more truthful responses;
NoVo~\cite{ho2025novo} achieves strong TruthfulQA accuracy through
majority voting over attention-head norms; and MIND~\cite{su2024mind}
and INSIDE~\cite{chen2024inside} show that internal states support
effective post-hoc hallucination detection. These contributions reveal
that internal model states carry actionable information about
truthfulness, and they leverage this information either to detect
hallucinations after generation or to steer activations during
inference. Inspired by these findings, we explore a third application:
using internal signals as a training-time reward, so that the
information they expose can shape policy learning itself.

\subsection{Experimental Details}
\label{Experimental_Details}
\subsubsection{Datasets Details}
\label{Datasets_Details}
We evaluate on three knowledge-intensive QA benchmarks spanning single-hop and multi-hop reasoning tasks.
\textbf{RAGTruth}~\citep{niu2024ragtruth},
\textbf{HotpotQA}~\citep{yang2018hotpotqa}, and
\textbf{2WikiMultiHopQA}~\citep{ho2020constructing}. RAGTruth focuses on
faithfulness in retrieval-augmented generation, while HotpotQA and
2WikiMultiHopQA require multi-hop reasoning over supporting evidence.
We use a fixed-evidence setting in which all methods receive the same
evidence pool. This controlled setup isolates the effect of policy
learning from retriever variability. Below we introduce the datesets in more detail:

\textbf{RAGTruth}~\citep{niu2024ragtruth} is a hallucination-focused corpus containing nearly 18,000 response-level annotations from diverse LLMs in standard RAG settings. We use its QA subset, which provides question--evidence--answer triples with explicit retrieval contexts. Each instance includes word-level hallucination annotations, making it particularly suitable for evaluating faithfulness.

\textbf{HotpotQA}~\citep{yang2018hotpotqa} contains 113K Wikipedia-based question-answer pairs requiring multi-hop reasoning over multiple supporting documents. Questions are diverse and not constrained to pre-existing knowledge schemas, with sentence-level supporting facts provided for explainability. We use the distractor setting where models must identify relevant information among 10 candidate paragraphs.

\textbf{2WikiMultiHopQA}~\citep{ho2020constructing} is a multi-hop dataset that combines structured (Wikidata) and unstructured (Wikipedia) sources. It contains approximately 190K examples with explicit reasoning paths expressed as entity-relation triples, enabling fine-grained evaluation of reasoning steps. Questions span four types: comparison, inference, compositional, and bridge-comparison.

To isolate our analysis from retriever quality and focus on the generation and reasoning components, we adopt a fixed-evidence setting where dataset-provided contexts serve as the evidence pool.





\subsubsection{Implementation Details}
\label{sec:implementation_details}
We instantiate the framework with three backbones: Qwen3-8B, Llama3-8B,
and Llama2-13B. The backbone weights stay frozen throughout training, and
we train three role-specific LoRA adapters for query planning, evidence
selection, and answer generation. Each adapter uses rank $16$ and scaling
factor $32$ on the query, key, value, and output projections.

\paragraph{Training stages.}
Training proceeds in two stages. We first warm-start the three adapters
with supervised fine-tuning for 1 epoch at learning rate
$2\times10^{-5}$. We then run PPO for 1 epoch at learning rate
$5\times10^{-6}$, with clip ratio $\epsilon=0.2$, KL coefficient
$\rho=0.1$, task reward weight $1.0$, and hallucination penalty weight
$\lambda=0.08$.

\paragraph{Hallucination critic.}
For each backbone, the critic reads attention-head norms from a fixed
window of intermediate layers and keeps the truth-correlated heads above
a $0.90$ quantile threshold. Each selected head contributes four
token-level statistics (mean, max, standard deviation, minimum), and a
small MLP maps these features to a scalar risk score. The critic is
trained with binary cross-entropy on TruthfulQA pairs and frozen
thereafter.

\subsubsection{Evaluation metrics.}
For HotpotQA and 2WikiMultiHopQA, we report Accuracy (Acc), Exact Match
(EM), and F1, following standard QA evaluation protocols. For RAGTruth,
we report Recall, EM, and F1. We use Recall instead of Accuracy as the
first RAGTruth metric because our evaluation focuses on whether
answer-relevant evidence tokens are recovered in the generated response.
Accordingly, we use ``R/A'' in averaged columns to denote Recall for
RAGTruth and Accuracy for HotpotQA and 2WikiMultiHopQA.

We additionally report \textbf{Answer Hallucination Rate (AHR)} on
RAGTruth using GPT-4o-mini as an automated judge. AHR measures the
fraction of generated answers that contain claims unsupported by the
provided evidence:
\begin{equation}
\mathrm{AHR}
=
\frac{
|\{a \in \mathcal{A}: \mathrm{contains\_hallucination}(a,e)\}|
}{
|\mathcal{A}|
},
\end{equation}
where $\mathcal{A}$ is the set of generated answers and $e$ denotes the
corresponding evidence. Unlike Recall, EM, or F1, AHR captures
unsupported additions that may be missed by surface-form metrics when an
answer contains the reference span but also introduces fabricated
details.

\subsubsection{Details about Baseline Models}
\label{sec:baseline_details}
We compare against methods spanning vanilla RAG, self-reflective approaches, and multi-agent systems:

\begin{itemize}[leftmargin=*,noitemsep,topsep=2pt]
    \item \textbf{Zeroshot w/ SFT}: The backbone LM fine-tuned on task data without retrieval augmentation.
    \item \textbf{Vanilla RAG}: Standard retrieve-then-generate pipeline using top-$k$ retrieved passages as context.
    \item \textbf{Vanilla RAG w/ SFT}: Vanilla RAG with additional supervised fine-tuning on RAG-formatted examples.
    \item \textbf{SelfRAG}~\citep{asai2024selfrag}: A self-reflective framework that trains LMs to adaptively retrieve on-demand and critique generations using learned reflection tokens, enabling controllable trade-offs between retrieval frequency and output quality.
    \item \textbf{MAIN-RAG}~\citep{chang-etal-2025-mainrag}: A multi-agent pipeline with independent query rewriting and answer generation modules, optimized separately via supervised learning.
    \item \textbf{RAG-DDR}~\citep{li2024ragddr}: A differentiable data rewards method that end-to-end trains RAG systems by aligning data preferences between retrieval and generation modules through rollout-based reward propagation.
    \item \textbf{MA-RAG}~\citep{nguyen2025marag}: A multi-agent RAG framework using collaborative chain-of-thought reasoning with specialized agents for planning, extraction, and answer synthesis.
    \item \textbf{MMOA-RAG}~\citep{chen2026mmoa}: A multi-agent RL approach that jointly optimizes query rewriter, selector, and generator modules using MAPPO with F1-based task rewards, representing the closest baseline to our method but without faithfulness-aware signals.
\end{itemize}

\subsection{Additional Results}
\subsubsection{Ablation Study}

\begin{table*}[tb]
  \centering
  \small
  \setlength{\tabcolsep}{2.5pt}
  \newcommand{\mc}[1]{\makebox[2.8em][c]{#1}}
  \begin{tabular}{lcccccccccccc}
    \toprule
    \multirow{2}{*}{Variant} &
    \multicolumn{3}{c}{RAGTruth} &
    \multicolumn{3}{c}{HotpotQA} &
    \multicolumn{3}{c}{2WikiMultiHopQA} &
    \multicolumn{3}{c}{Average} \\
    \cmidrule(lr){2-4}
    \cmidrule(lr){5-7}
    \cmidrule(lr){8-10}
    \cmidrule(lr){11-13}
    & Recall & EM & F1
    & Acc & EM & F1
    & Acc & EM & F1
    & R/A & EM & F1 \\
    \midrule
    w/o critic
      & \mc{43.82} & \mc{35.96} & \mc{44.42}
      & \mc{47.26} & \mc{36.75} & \mc{45.96}
      & \mc{47.04} & \mc{43.35} & \mc{48.83}
      & \mc{46.04} & \mc{38.69} & \mc{46.40} \\
    w/o Q update
      & \mc{44.70} & \mc{36.84} & \mc{45.46}
      & \mc{47.19} & \mc{38.41} & \mc{47.02}
      & \mc{48.93} & \mc{44.20} & \mc{49.87}
      & \mc{46.94} & \mc{39.82} & \mc{47.45} \\
    w/o E update
      & \mc{47.18} & \mc{37.31} & \mc{46.11}
      & \mc{47.72} & \mc{38.89} & \mc{47.63}
      & \mc{49.45} & \mc{44.67} & \mc{50.48}
      & \mc{48.12} & \mc{40.29} & \mc{48.07} \\
    w/o G update
      & \mc{43.91} & \mc{35.32} & \mc{43.46}
      & \mc{45.38} & \mc{34.62} & \mc{44.02}
      & \mc{46.06} & \mc{41.58} & \mc{46.91}
      & \mc{45.12} & \mc{37.17} & \mc{44.80} \\
    \midrule
    \rowcolor{myteal!30}
    Full
      & \mc{\textbf{48.08}} & \mc{\textbf{38.18}} & \mc{\textbf{47.89}}
      & \mc{\textbf{49.34}} & \mc{\textbf{40.02}} & \mc{\textbf{48.44}}
      & \mc{\textbf{50.62}} & \mc{\textbf{45.84}} & \mc{\textbf{50.63}}
      & \mc{\textbf{49.35}} & \mc{\textbf{41.35}} & \mc{\textbf{48.99}} \\
    \bottomrule
  \end{tabular}
  \caption{Ablation results with Qwen3-8B. In agent-update ablations,
  the named agent remains in the Q--E--G inference pipeline but is not
  updated during PPO. The average R/A column follows
  Table~\ref{tab:main_results}: Recall for RAGTruth and Accuracy for
  HotpotQA and 2WikiMultiHopQA.}
  \label{tab:ablation}
\end{table*}

We provide comprehensive ablation analysis across all datasets and model architectures to validate each component's contribution. Table~\ref{tab:ablation} presents the complete ablation results on Qwen3-8B.

\subsubsection{Additional Critic Transfer Diagnostics}
\label{sec:critic_transfer_appendix}

This section provides additional diagnostics for the TruthfulQA-calibrated
critic used in our hallucination-aware reward. The goal is not to train a
task-specific hallucination classifier for each RAG benchmark, but to
verify whether a critic calibrated on general truthful/false answer pairs
can provide a useful auxiliary signal during downstream RAG policy
optimization.

\paragraph{Truthfulness calibration.}

We first evaluate the critic on held-out TruthfulQA truthful/false answer
pairs. After searching over the head-selection configurations, the best
critic achieves \textbf{75.92\% AUROC}, indicating that the selected
head-norm features contain a substantive truthfulness signal. The critic
is then kept frozen and used only as a scalar reward-shaping term during
RAG policy optimization.

To examine whether the same internal features also capture evidence faithfulness, we train an evidence-aware critic using RAGTruth's hallucination labels. Using the same attention-head-norm features, the critic achieves 74.05\% AUROC on held-out RAGTruth data. This result shows that the internal features contain information related to evidence support in addition to general truthfulness.

We also examine whether the critic score is driven by simple surface properties. Its correlation with answer length is $-0.1103$, while its correlation with answer--evidence word overlap is $-0.1077$. Both values are close to zero, suggesting that the critic score is not simply a proxy for answer length or lexical overlap.

\paragraph{Cross Judge Robustness}
\label{sec:cross_judge_robustness}

To examine whether the Answer Hallucination Rate (AHR) depends on a
particular automatic judge, we repeat the evaluation using Qwen2.5-7B,
a model from a different family from GPT-4o-mini. Table~\ref{tab:cross_judge}
compares the AHR scores produced by the two judges on two backbones.

\begin{table}[t]
    \centering
    \small
    \setlength{\tabcolsep}{6pt}
    \begin{tabular}{lcc}
        \toprule
        Backbone & GPT-4o-mini & Qwen2.5-7B \\
        \midrule
        Qwen3-8B   & 21.24 & 21.11 \\
        Llama2-13B & 27.69 & 24.12 \\
        \bottomrule
    \end{tabular}
    \caption{Answer Hallucination Rate (\%, lower is better) evaluated
    using two independent automatic judges.}
    \label{tab:cross_judge}
\end{table}

The two judges produce similar AHR scores on both backbones, with the
overall trend remaining consistent across model families. This suggests
that the reported hallucination rates are not specific to GPT-4o-mini.

\paragraph{Shuffled-score control.}
To test whether downstream gains come from the sample-level critic signal
rather than from the marginal distribution of penalty values, we run a
shuffled-score control. In this control, the same frozen critic computes
a risk score $H$ for each rollout, but the scores are randomly reassigned
across samples within each PPO update batch before computing the reward:
\begin{equation}
R = \alpha R_{\mathrm{task}} - \lambda H_{\mathrm{shuffled}}.
\end{equation}
This preserves the scale and distribution of critic penalties while
breaking their correspondence with the specific generated answers. If the
critic were merely acting as a generic regularizer, this control would be
expected to perform similarly to the calibrated critic. Instead, as shown
in Table~\ref{tab:critic_transfer_diag}, the calibrated critic
outperforms the shuffled-score control, suggesting that the sample-level
risk signal is informative.

\paragraph{Reward-weight calibration.}
We also vary the hallucination penalty weight $\lambda$ to examine the
trade-off between the task reward and the internal risk penalty. Among the tested settings, $\lambda=0.08$ gives the strongest RAGTruth
Recall, EM, and F1, while the shuffled-score control performs worse than
the calibrated critic. This suggests that both the scale of the penalty
and its sample-level correspondence with generated answers matter.
\begin{table}[t]
    \centering
    \small
    \setlength{\tabcolsep}{5pt}
    \begin{tabular}{lcccc}
        \toprule
        Setting & $\lambda$ & Recall & EM & F1 \\
        \midrule
        w/o critic              & 0.00 & 43.82 & 35.96 & 44.42 \\
        shuffled-score critic   & 0.08 & 38.20 & 37.50 & 46.21 \\
        critic                  & 0.04 & 47.98 & 35.03 & 46.96 \\
        critic                  & 0.08 & 48.08 & 38.18 & 47.89 \\
        \bottomrule
    \end{tabular}
    \caption{Additional RAGTruth diagnostics for the TruthfulQA-calibrated
    critic. The shuffled-score control preserves the marginal distribution
    of critic penalties but breaks their sample-level correspondence with
    rollouts. The results show that the critic signal is useful as reward
    shaping, while the penalty weight affects the balance between the task reward and the internal risk penalty.}
    \label{tab:critic_transfer_diag}
\end{table}

These diagnostics support two conclusions. First, the critic captures a
real internal truthfulness signal on TruthfulQA, as reflected by its
held-out AUROC. Second, when transferred to RAG policy optimization, the
signal is most useful as a calibrated reward-shaping term: preserving the
sample-level critic scores matters, but the penalty weight must be
balanced against the task reward. This matches the design of our method,
which treats the critic as a weak complementary signal rather than a
standalone judge of RAG hallucination.

\subsection{Prompt Templates}
\label{sec:prompt_templates}

Each agent receives a fixed system prompt and a structured user prompt
filled at episode time. The templates are shared across all backbones
and datasets. Figure~\ref{fig:prompt_templates} shows the complete
prompts for all three agents and the LLM-as-a-judge faithfulness
evaluator. For Qwen3-8B, agent prompts are wrapped with the model's
chat template and thinking mode is disabled
(\texttt{enable\_thinking=False}) to prevent \texttt{<think>} output
from interfering with answer parsing.

\begin{figure*}[t]
\centering
\small

\begin{tcolorbox}[
  colback=white,
  colframe=black!65,
  colbacktitle=black!75,
  coltitle=white,
  fonttitle=\bfseries\small,
  title={Agent Q \textemdash{} Query Planner},
  sharp corners,
  boxrule=0.6pt,
  left=5pt, right=5pt, top=4pt, bottom=4pt,
  width=\linewidth
]
\textbf{System:} You are a professional assistant skilled at rewriting
complex or unclear questions into simpler, more searchable
sub-questions. The rewritten sub-questions must have logical connections
and dependencies, without being overly repetitive in meaning. Avoid
using vague demonstrative pronouns.

\medskip
\textbf{User:} Please rewrite or decompose the following question into
sub-questions, making them easier to search for answers in a search
engine.

\smallskip
Original question: \textit{\{question\}}

\smallskip
Requirements: (1)~Output one sub-question per line. (2)~Avoid duplicate
sub-questions. (3)~Avoid vague demonstrative pronouns.
(4)~At most 4 sub-questions.

\smallskip
Output the sub-questions only, without any additional explanation.
\end{tcolorbox}

\begin{tcolorbox}[
  colback=white,
  colframe=black!65,
  colbacktitle=black!75,
  coltitle=white,
  fonttitle=\bfseries\small,
  title={Agent E \textemdash{} Evidence Selector},
  sharp corners,
  boxrule=0.6pt,
  left=5pt, right=5pt, top=4pt, bottom=4pt,
  width=\linewidth
]
\textbf{System:} You are a helpful assistant. Your task is to select
the most relevant documents from the candidate list that can help answer
the question. Output only the document IDs.

\medskip
\textbf{User:} Question: \textit{\{question\}}

\smallskip
Candidate Documents: \textit{\{documents\}}

\smallskip
Select the documents most helpful for answering the question.
Output format: \texttt{Document0,Document2,Document5}
(comma-separated IDs). Only output the selected document IDs, nothing
else.
\end{tcolorbox}

\begin{tcolorbox}[
  colback=white,
  colframe=black!65,
  colbacktitle=black!75,
  coltitle=white,
  fonttitle=\bfseries\small,
  title={Agent G \textemdash{} Answer Generator},
  sharp corners,
  boxrule=0.6pt,
  left=5pt, right=5pt, top=4pt, bottom=4pt,
  width=\linewidth
]
\textbf{System:} You are a helpful, respectful and honest assistant.
Your task is to predict the answer to the question based on the given
documents. If you do not know the answer, please do not share false
information. Answer the question as accurately and briefly as possible.

\medskip
\textbf{User:} Question: \textit{\{question\}}

\smallskip
Reference Documents: \textit{\{documents\}}

\smallskip
Based on the above documents, answer the question as briefly and
accurately as possible. Only give the brief and accurate answer, nothing
else.
\end{tcolorbox}

\begin{tcolorbox}[
  colback=white,
  colframe=black!65,
  colbacktitle=black!75,
  coltitle=white,
  fonttitle=\bfseries\small,
  title={LLM Judge \textemdash{} Faithfulness Evaluator},
  sharp corners,
  boxrule=0.6pt,
  left=5pt, right=5pt, top=4pt, bottom=4pt,
  width=\linewidth
]
\textbf{System:} You are an evidence-grounded evaluator. Given a
Question, Evidence passages, and a Model Answer, determine whether the
Model Answer is fully supported by the Evidence \emph{only}.
Use ONLY the provided Evidence; do not use external knowledge or
common sense. Do not infer missing details. If any key claim is not
directly supported or is contradicted, label \textbf{UNSUPPORTED}.
Quote short verbatim spans from the passages you relied on.
Return a JSON verdict with fields: \texttt{label}
(\texttt{SUPPORTED}/\texttt{UNSUPPORTED}), \texttt{issue\_type},
\texttt{confidence}, \texttt{rationale}, and \texttt{evidence\_spans}.

\medskip
\textbf{Input:} Question: \textit{\{question\}}

\smallskip
Evidence: \texttt{[passage\_1]} \textit{\{text\}} $\;\cdots$

\smallskip
Model Answer: \textit{\{answer\}}

\smallskip
Output JSON:
\end{tcolorbox}

\caption{Prompt templates for the three RAG agents (Q, E, G) and the
LLM-as-a-judge faithfulness evaluator. Slots in italics
(\textit{\{question\}}, \textit{\{documents\}}, \textit{\{answer\}},
\textit{\{text\}}) are filled at inference time. Agent~E additionally
receives the planned sub-questions from Agent~Q as a soft conditioning
signal. The judge outputs a structured JSON verdict used in the
Faithfulness Analysis (Section~\ref{sec:faithfulness}).}
\label{fig:prompt_templates}
\end{figure*}

\subsection{Case Study}
\label{sec:case_study}

Figure~\ref{fig:case_study} shows two representative inference traces
illustrating the three-agent pipeline and the hallucination critic.
Example 1 is a multi-hop question from HotpotQA where Agent~E
successfully narrows the retrieved passages to the two relevant ones,
enabling Agent~G to produce a correct and grounded answer. Example~2 is
a causality question from 2WikiMultiHopQA where the retrieved passages
do not contain the required information; the critic correctly identifies
the resulting output as problematic and applies a training penalty.

\begin{figure*}[t]
\centering
\small

\begin{tcolorbox}[
  colback=white,
  colframe=black!65,
  colbacktitle=black!75,
  coltitle=white,
  fonttitle=\bfseries\small,
  title={Example 1 \textemdash{} Correct pipeline (HotpotQA)},
  sharp corners,
  boxrule=0.6pt,
  left=5pt, right=5pt, top=4pt, bottom=4pt,
  width=\linewidth
]
\textbf{Question:}
\textit{The football manager who recruited David Beckham managed
Manchester United during what timeframe?}

\medskip
\textbf{Agent~Q:} The question is well-formed; passed to retrieval
unchanged.

\smallskip
\textbf{Agent~E:} Selects two relevant passages (Alex Ferguson's
biography and Manchester United's managerial history) from 10
retrieved candidates; discards the remaining 8.

\smallskip
\textbf{Agent~G:} \textit{``1986 to 2013''} \quad (correct)

\smallskip
\textbf{Critic:} Low hallucination score \quad $\Rightarrow$ reward
unpenalised.
\end{tcolorbox}

\smallskip

\begin{tcolorbox}[
  colback=white,
  colframe=black!65,
  colbacktitle=black!75,
  coltitle=white,
  fonttitle=\bfseries\small,
  title={Example 2 \textemdash{} Critic detects problematic output (2WikiMultiHopQA)},
  sharp corners,
  boxrule=0.6pt,
  left=5pt, right=5pt, top=4pt, bottom=4pt,
  width=\linewidth
]
\textbf{Question:}
\textit{Why did Grand Duke Kirill Vladimirovich of Russia's wife die?}

\medskip
\textbf{Agent~E:} Selects passages about the Grand Duke; none contain
an explicit cause of death.

\smallskip
\textbf{Agent~G:} \textit{``not mentioned''} \quad (incorrect; gold:
\textit{``stroke''})

\smallskip
\textbf{Critic:} High hallucination score \quad $\Rightarrow$
hallucination penalty applied.
\end{tcolorbox}

\caption{Two inference traces illustrating the pipeline and the
hallucination critic. In Example~1, Agent~E focuses retrieval to two
relevant passages and Agent~G produces a correct answer confirmed by a
low critic score. In Example~2, the critic flags a refusal-style output
caused by insufficient evidence, applying a training penalty.}
\label{fig:case_study}
\end{figure*}

\subsection{Usage Claim of AI assistants}

We use LLM for grammar and spelling checks only, with prompt ``Proofread the sentences''. All conceptual development, analysis, writing, and editing were carried out solely by the authors without LLM assistance.

\end{document}